\documentclass[aps,twocolumn,nofootinbib,preprintnumbers,superscriptaddress,citeautoscript]{revtex4-1}
\pdfoutput=1
\usepackage{graphicx}
\usepackage{epsfig,amsmath,amsthm,amsfonts,mathrsfs,amssymb,mathtools}
\usepackage[usenames,dvipsnames,table]{xcolor}
\usepackage{hyperref}
\usepackage[normalem]{ulem}
\allowdisplaybreaks

\newcommand{\EW}{\mathrm{EW}}
\newcommand{\OW}{\mathrm{OW}}
\newcommand{\W}{\mathrm{W}}
\DeclarePairedDelimiter{\abs}{\lvert}{\rvert}

\DeclareRobustCommand{\SkipTocEntry}[5]{}

\hypersetup{
    pdfstartview={FitH},    
    pdftitle={},    
    pdfauthor={Massimiliano Rota, Sean J. Weinberg},     
    colorlinks=true,       
    linkcolor=blue,          
    citecolor=red,        
    filecolor=magenta,      
    urlcolor=blue           
}

\definecolor{green}{rgb}{0.1,0.8,0.2}
\definecolor{myblue}{rgb}{0.12,0.51,0.88}

\theoremstyle{definition}
\newtheorem{theorem}{Theorem}

\def\bra#1{\langle #1 |}
\def\ket#1{| #1 \rangle}

\begin{document}

\title{Holographic Shannon Entropy: \\ The Outer Entropy of Entanglement Wedges}


\author{Sean J. Weinberg}
\email{sjasonw@physics.ucsb.edu}
\affiliation{Department of Physics, University of California, Santa Barbara, CA 93106, USA}

\begin{abstract}{
We introduce a simple geometrical construction similar to covariant holographic entanglement entropy but with
the addition of a new term proportional to boundary region volume.  This new procedure has properties strongly resembling 
classical Shannon entropy of probability distributions rather than von Neumann entropy, so we
call the quantity holographic Shannon entropy.  The holographic Shannon entropy of
a region $A$ is divergent in AdS/CFT, but upon regulation, it appears to be equal to the outer entropy
of the entanglement wedge of $A$ with the entanglement wedge of the complement of $A$ held fixed.  The construction is unambiguous when applied to compact surfaces
with convex shape in general spacetimes obeying the null curvature condition (of which AdS with a regulator is a special case).
In this context we prove
that holographic Shannon entropy obeys monotonicity, a key property of Shannon entropy, as well as 
all known balanced inequalities of dynamical holographic entanglement entropy.  In the static case, we explain
why there must exist some classical probability distribution on random variables locally distributed on the boundary
with the property that the Shannon entropies of all marginals are exactly reproduced by the holographic Shannon
entropy formula.
}
\end{abstract}

\maketitle

\section{Overview}

Perhaps the most significant recent development to arise from research on holographic entanglement entropy \cite{RT, HRT} was establishing
the modern form of subregion duality \cite{DHW}.  The possibility that a region $A$ on the boundary of an asymptotically AdS spacetime is dual to a specific bulk subregion
was originally investigated from a causal structure perspective \cite{subregion_old}, but it has since been discovered that the entanglement wedge  of $A$ \cite{entanglement_wedge}, the bulk region between $A$ and its entangling surface, is the most reasonable thing to call the bulk region associated with $A$.
Subregion duality is a major breakthrough.  It reveals the role of quantum error correction in quantum gravity \cite{error_correction_adscft} and allows for some notion
of spatial subsystems in quantum gravity, a subject where locality is notoriously absent \cite{DG_1,DG_2}.

Subregion duality leads to an intriguing although poorly formed question: how many quantum states fit inside the entanglement wedge of a boundary region $A$?
Without worrying about the precise meaning of this question, we can give two distinct answers.  $A$ is a subregion on which a CFT is defined, and upon fixing
a UV regulator, the region has a number of degrees of freedom proportional to its volume $|A|$ so $e^{|A|/ 4 G_{\mathrm N}}$ is a potential guess.
On the other hand, a somewhat different answer is suggested by the Bousso bound (also known as the covariant entropy bound) \cite{Bousso_bound}.
 If $A$ is regulated in a way that makes its ingoing future null expansion non-positive, then the extremality of the HRT surface of $A$
 allows us to apply the Bousso bound to the spacelike surface $A \cup \gamma(A)$ where $\gamma(A)$ is the HRT surface of $A$.
 The upper bound on light sheet entropy is
the sum of the CFT volume of $A$ plus the area of $\gamma(A)$.  The former has a divergence that dominates over the
latter, but perhaps we are losing a great deal of fascinating structure by ignoring the interplay between the two terms.

\begin{figure}
  \includegraphics[width=9cm]{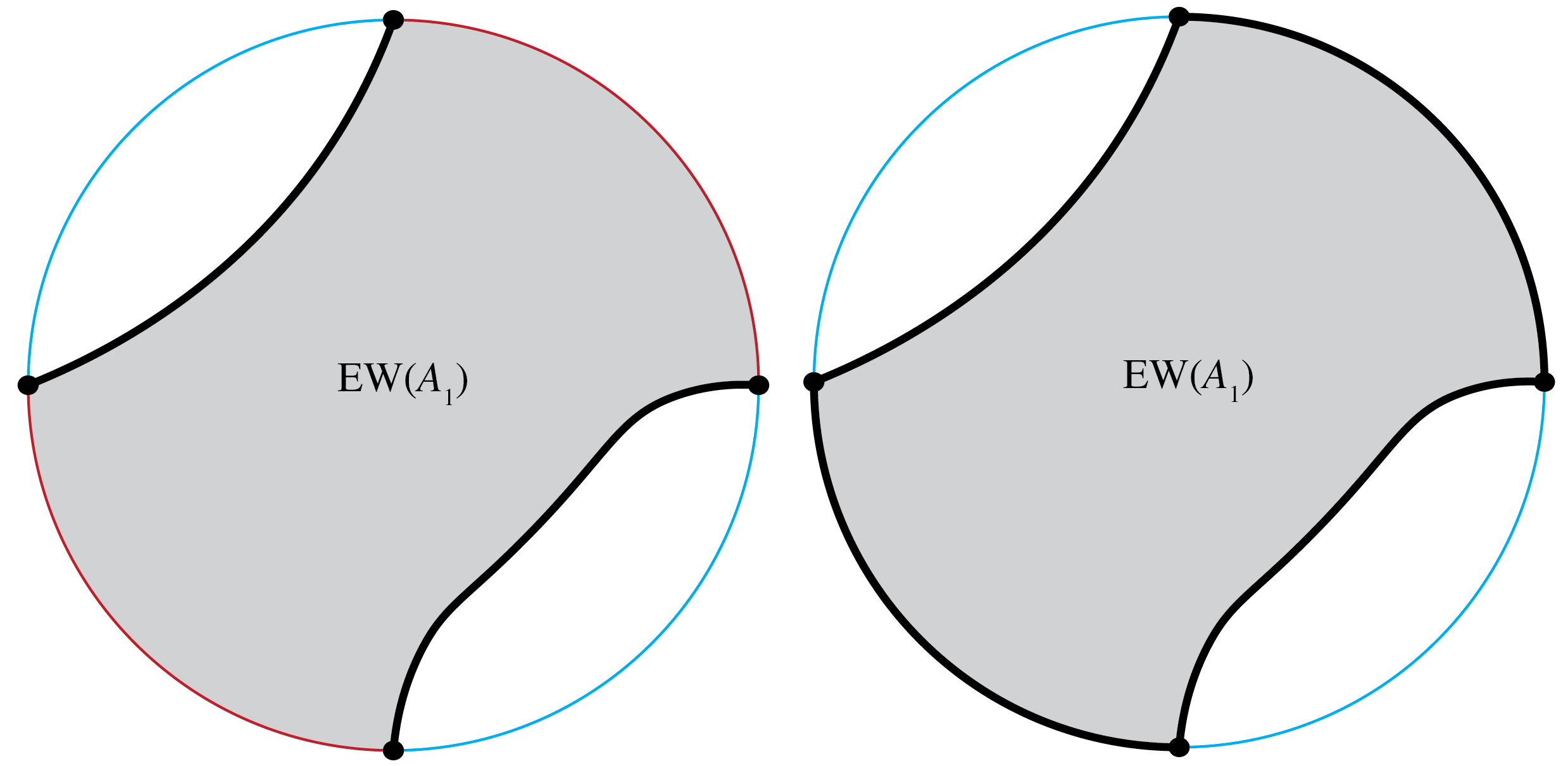}
  \caption{Holographic Shannon definition is defined almost identically to the HRT formula for covariant holographic entanglement entropy.
  In the figure on the left above, the area of the black surface gives the usual von Neumann entropy of the red region $A_1$.  In the figure on the
  right, the area of the black surface, which now includes $A_1$ as well as its entangling surface, produces what we call holographic Shannon entropy.}
  \label{fig:definition}
\end{figure}

In this paper, we choose to not ignore the second or the first term that arises from the Bousso bound when applied to an entanglement wedge.
In doing so, we present a geometrical construction with significance that overlaps between quantum gravity and information theory.  Our construction
is very simple: we compute the quantity
\begin{equation}
\label{basic_def}
H(A) = \frac{1}{4 G_{\mathrm N}} \left( |A| + |\gamma(A)| \right)
\end{equation}
where $|A|$ is the regulated area of a boundary region $A$ and $|\gamma(A)|$ is the regulated area of the the HRT surface of $A$.  
Figure \ref{fig:definition} illustrates the procedure.
Equation \eqref{basic_def} is much cleaner when applied to convex compact surfaces in general spacetimes obeying the null curvature condition
along the lines of proposals for extending holographic entanglement to general spacetimes \cite{surface_state,SW} in which case the first term is not much larger than the second.

The function $H$ in equation \eqref{basic_def} will be referred to as \emph{holographic Shannon entropy} regardless of whether it is applied
in AdS/CFT with a cutoff or on a convex surface in a non-AdS spacetime.  The motivation for calling it a Shannon entropy comes from a collection
of results.  In the case where the static RT formula applies for von Neumann entropy, a result due to Gross and Walter \cite{stabilizer_classical}
allows us to prove that there must exist a probability distribution $p$ on a collection of (correlated ) random variables locally distributed along the boundary
such that for any subregion $A$, the marginal probability distribution $p_A$ has Shannon entropy given precisely by \eqref{basic_def}.

In the dynamical case, we are not currently able to prove the existence of such a probability distribution, but we can come close.  
We prove that $H$ satisfies monotonicity: $H(AB) \geq H(A)$, a 
critical feature of Shannon entropy, and we also confirm that it satisfies every known balanced inequality for the dynamical holographic entropy cone.
This immediately implies \cite{Mm_not_enough} that $H$ satisfies a number of known inequalities required for Shannon entropy including the Zhang-Yeung
inequality \cite{zhang_yeung}.
  If the dynamical
holographic entropy cone turns out to be equal to the static entropy cone, a very important open problem\cite{Mm_not_enough}, then we 
can can conclude that a probability distribution realizes $H$ even in the general dynamical case.

Showing that $H$ is a Shannon entropy is an important step, but it raises more questions then it answers.  The probability distribution
on the boundary seems to be a classical analog of a density matrix in AdS/CFT, but it is quite unclear what the meaning of
the distribution is.  This paper does not give a unique or physically motivated definition of the probability distribution,
we only show that it exists in many cases.

Fortunately, holographic Shannon entropy itself stands on stronger physical grounds.  In addition to our comments above about the
covariant entropy bound, the best hint about the underlying meaning of $H$ comes from an outer entropy
computation based on the ideas of \cite{outer_1, outer_2}.  Computing the outer entropy of a surface means holding the spacetime outside
of the surface fixed while coarse graining over all possibilities for the unknown region in the interior of the surface.  The most famous
result about outer entropy is that the outer entropy of an apparent horizon is equal to its area, but there is a quickly growing collection
of other intriguing results \cite{Bousso:2018fou,Nomura:2018aus,Engelhardt:2019btp}.

In the following section, we will make the case that the holographic Shannon entropy of a region $A$ is the regulated outer entropy of its entangling surface
with the complement entanglement wedge held fixed.  Our argument is plagued by the divergence of $H(A)$, but the result still appears to be the only reasonable
answer to give for the outer entropy of the entanglement wedge of $A$ upon regulation.  This is consistent with what we would speculate with the Bousso bound.
After all, the original purpose of holographic entropy bounds is to estimate an upper bound for the gravitational entropy inside of a surface, a counting of the number
of spacetimes that could be hidden behind a wall.  Our results add some confidence to this idea: the area of an entanglement wedge counts the number
of states consistent with the fixed geometry in the complement wedge.

\section{Motivation: outer entropy of the entanglement wedge}
\label{sec:motivation}
The geometrical construction of holographic Shannon entropy can be motivated in two separate ways.
The first is an attempt to answer a question related to gravitational entropy which we now pose.

Consider a classical asymptotically locally AdS spacetime $M_0$ in the context of AdS/CFT in the large $N$ limit with conformal boundary $\partial M_0$.  Let $B_0$ be a Cauchy surface for $\partial M_0$.
We are assuming that  $\partial M_0$ is the entire boundary so that there is a pure quantum state $\ket{\psi_0}$ on a CFT on $B_0$ which describes the spacetime.
Divide $B_0$ into two disjoint subregions: $A_1$ and  $A_2$ so that $B_0 = A_1 \cup A_2$.
Consider the entanglement wedge of $A_1$, denoted $\EW(A_1, M_0)$ or just $\EW(A_1)$ if we have clearly specified $M_0$.  
We also use the terminology \emph{outer wedge} of $A_1$, denoted $\OW(A_1, M_0)$, to refer to the entanglement wedge of $A_2$  rather
than $A_1$.

Suppose that we only have access to the outer wedge of $A_1$ and that the the entanglement wedge of $A_1$ is hidden from our observation.  
What entropy should be assigned to quantify this lack of information about the entire bulk?  In other words, \emph{what is the outer entropy of the entanglement
wedge of $A_1$}?

\begin{figure}
  \includegraphics[width=\linewidth]{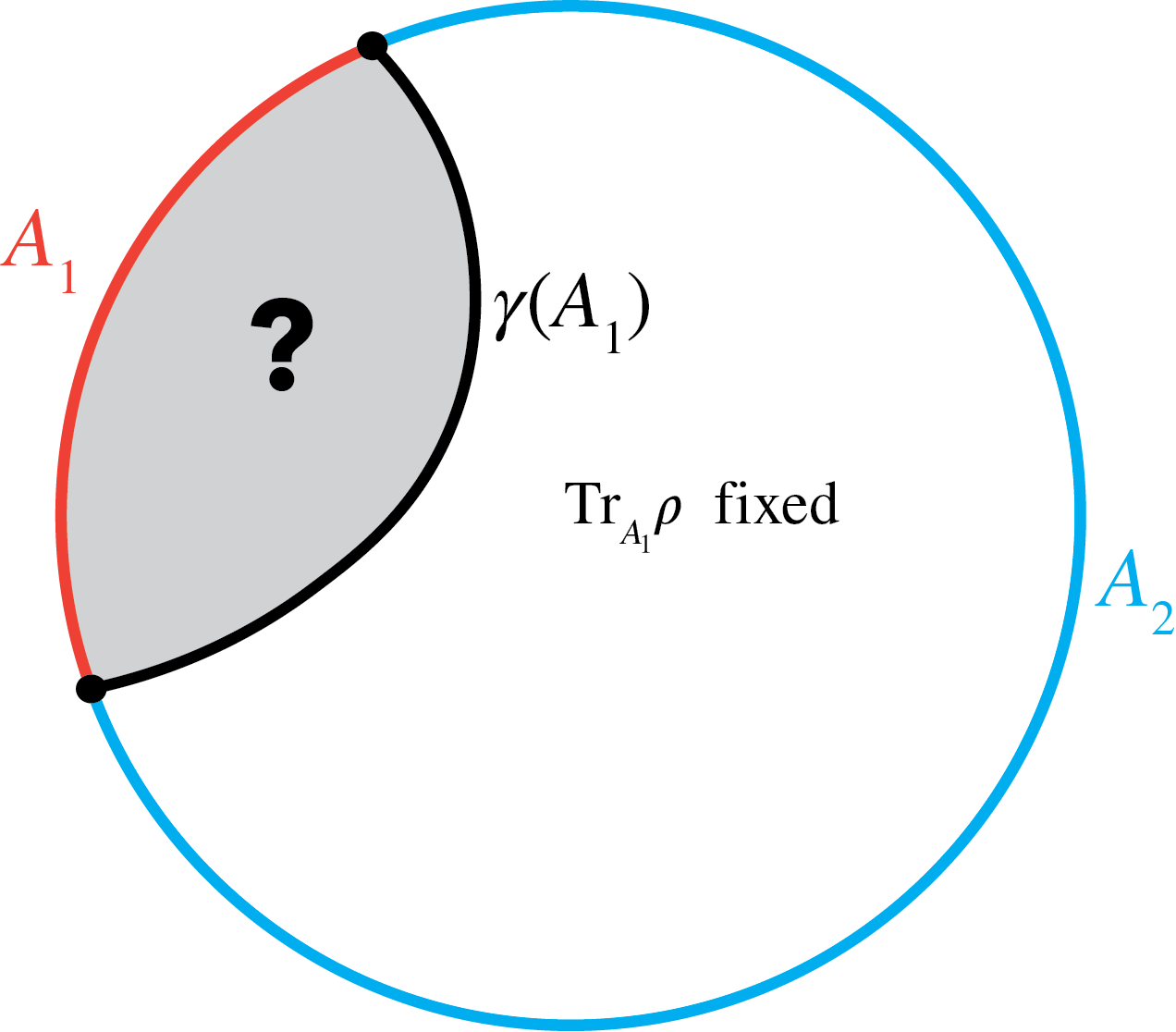}
  \caption{The collection $\Omega$ is the set of density matrices on $A_1 A_2 = B_0$ with the property that  ${\rm Tr}_{A_1} \rho$ is the same
  as the reduced density matrix in the known entanglement wedge of $A_2$.  $\Omega$ provides a way to quantify the entropy associated with the missing information about
  the wedge of $A_1$.}
  \label{fig:outer_density}
\end{figure}

Consider a collection of density matrices on the the Hilbert space of $B_0$ consistent with the outer wedge of $A_1$ being held fixed.  Recall that we had access to a specific
pure quantum state $\ket{\psi_0}$ which was associated with the spacetime $M_0$.  Put
\[
\Omega = \left\{ \rho \; {\rm on }\; B_0 \: \big| \: {\rm Tr}_{A_1} \left(\rho\right) = {\rm Tr}_{A_1} \left(\ket{\psi_0} \bra{\psi_0}\right) \right\}.
\]
In other words, $\Omega$ is the collection of all density matrices on $B_0$ subject to the condition that if $\rho \in \Omega$, then upon tracing out $A_1$ to a reduced 
$\rho_{A_2}$, we always find that $\rho_{A_2} =\rho^0_{A_2}$ where $\rho^0_{A_2} = \ket{\psi_0} \bra{\psi_0}$.  This concept is illustrated in figure \ref{fig:outer_density}.

We now attempt to compute a quantity we denote by $\overset{\circ}{S}(A_1)$ defined as
\begin{equation}
\label{maximization}
\overset{\circ}{S}(A_1) = \sup \left\{ S(\rho) \: \big| \: \rho \in \Omega \right\}
\end{equation}
where $S$ denotes von Neumann entropy.  This calculation should provide an estimate of the desired gravitational entropy of the entanglement wedge of $A_1$.
At heart it is a very similar calculation to that of \cite{outer_1,outer_2} because we are varying over density matrices consistent with a condition on the outer wedge $\OW(A_1, M_0)$.
The notation $\overset{\circ}{S}(A_1)$ is somewhat unclear because it depends not only on the region $A_1$ but also on the original reduced density matrix on $A_2$ as well
as a cutoff prescription on the CFT.

The quantity $\overset{\circ}{S}$ is divergent, but this is no reason to ignore its structure upon regulation. Fix some regularization procedure that makes it so that von Neumann
entropy cannot diverge.  More precisely, if $A$ is a subregion of $B_0$, we are now guaranteed that there is a (finite) positive number $S_\star(A)$ which is the maximum of
entropies on $A$ when varying over all density matrices on the Hilbert space.  We are now guaranteed by subadditivity that if $\rho \in \Omega$,
\begin{equation}
\label{upper_bound}
S(\rho) \leq S(\rho_{A_1}) + S(\rho_{A_2}) \leq S_\star(A_1) + \frac{|\gamma(A_1)|}{4 G \hbar}
\end{equation}
where we have replaced $S(\rho_{A_2})$ by the von Neumann entropy computed by the HRT formula for the region $A_2$ in the state $\psi_0$.  This is a justified manipulation
because $\rho \in \Omega$.  Note that the area of the HRT surface $\gamma(A_1)$ is cut off by the regularization, but it is still computed in the original bulk spacetime $M_0$
associated with the pure state $\psi_0$.

The upper bound \eqref{upper_bound} can be saturated because we can take a density matrix $\rho^\star_{A_1}$ with the maximal entropy $S_\star(A_1)$ on $A_1$
and put
\[
\rho = \rho^\star_{A_1} \otimes  {\rm Tr}_{A_1} \left(\ket{\psi_0} \bra{\psi_0}\right)
\]
which is obviously in $\Omega$ and obviously saturates \eqref{upper_bound}.  We thus have
\begin{equation}
\overset{\circ}{S}(A_1) = S_\star(A_1) + \frac{|\gamma(A_1)|}{4 G_{\mathrm N}}
\end{equation}
which is precisely our formula for holographic Shannon entropy if we interpret $S_\star(A_1)$ as the cutoff area (or CFT volume) of $A_1$.

\subsection*{Geometrical outer entropy}

This section can be skipped without a major loss of continuity for readers more interested in the properties of holographic Shannon entropy
rather than its ``derivation.''

There another very similar way to arrive at the holographic Shannon entropy formula which may appeal more to a relativist because it 
is a purely geometrical construction that does not rely on statements about density operators.  This second approach is directly
based on the concept of outer entropy.  However, we will need to generalize the definition of outer entropy in a way that may cause some
discomfort and to be as clear as possible about the strengths and weaknesses of this approach, we use the term \emph{geometrical
outer entropy} below to distinguish from the original definition of outer entropy given in \cite{outer_1}.  

The situation is similar to the above: $M_0$ is a fixed asymptotically locally AdS spacetime with $D$ dimensions that satisfies the null curvature condition.  The conformal boundary $\partial M_0$
may consist of multiple connected components.
$B_0$ is a ($D-2$ dimensional) Cauchy surface of $\partial M_0$.  Note that the number of connected components of $B_0$ is the same as the number of components of $\partial M_0$. 
Let $\Sigma$ be a spacelike AdS-Cauchy surface for $M_0$ with $B_0 \subset \Sigma$.  Let $\sigma$ be a $D-2$ dimensional (spacelike) submanifold of $\Sigma$.  $\sigma$ will then
be homologous to some subregion of $B_0$ which may or may not be all of $B_0$.  In any case, we find $A_1, A_2 \subset B_0$ with $\sigma$ anchored to $A_1 \cap A_2$, $A_1 \cup A_2 = B_0$,
and such that the regions only intersect at their mutual boundaries.  This is the standard setup for computing holographic entanglement entropy of $A_1$ in the case where $\sigma$ is extremal, but
we are not making that assumption yet.  $\sigma$ need not have anything to do with an entanglement wedge.
 
Let $K$ denote the subset of $\Sigma$ that lies between $\sigma$ and $A_2$.  The domain of dependence of $K$ is a dimension $D$ region which is called the \emph{outer wedge} of $\sigma$
on the $A_2$ side.  We will denote it simply by $\W(\sigma, A_2)$.  We will \emph{not} call this the outer wedge of $A_1$ in the spacetime $M_0$ because $\sigma$ is not an HRT surface. 
Our insistence to include $A_2$ as an argument to $\W$ is because $\sigma$ is not necessarily homologous to the boundary.

We now consider the collection of all spacetimes that are identical to $M_0$ in $\W(\sigma, A_2)$ but otherwise unconstrained. 
Suppose that $\bar{M}$ is an asymptotically locally AdS spacetime obeying the null curvature and generic condition with the property that
the conformal boundary of $\bar{M}$ has copies $\bar{A_1}$ and  $\bar{A_2}$ of $A_1$ and $A_2$ respectively.  We will
say that $\bar{M}$ is \emph{equivalent to $M_0$ on the $A_2$ side of $\sigma$} if there
exists a dimension $D$ subregion $W$ of $\bar{M}$ with $D \cap \partial \bar{M} = \bar{A_2}$ such that $W$ is isometric to
the outer wedge $\W(\sigma, A_2)$.
The collection of all spacetimes equivalent to $M_0$ on the $A_2$ side of $\sigma$ is denoted by $\mathcal{C}(M_0,\sigma, A_2)$.

\begin{figure}
  \includegraphics[width=9cm]{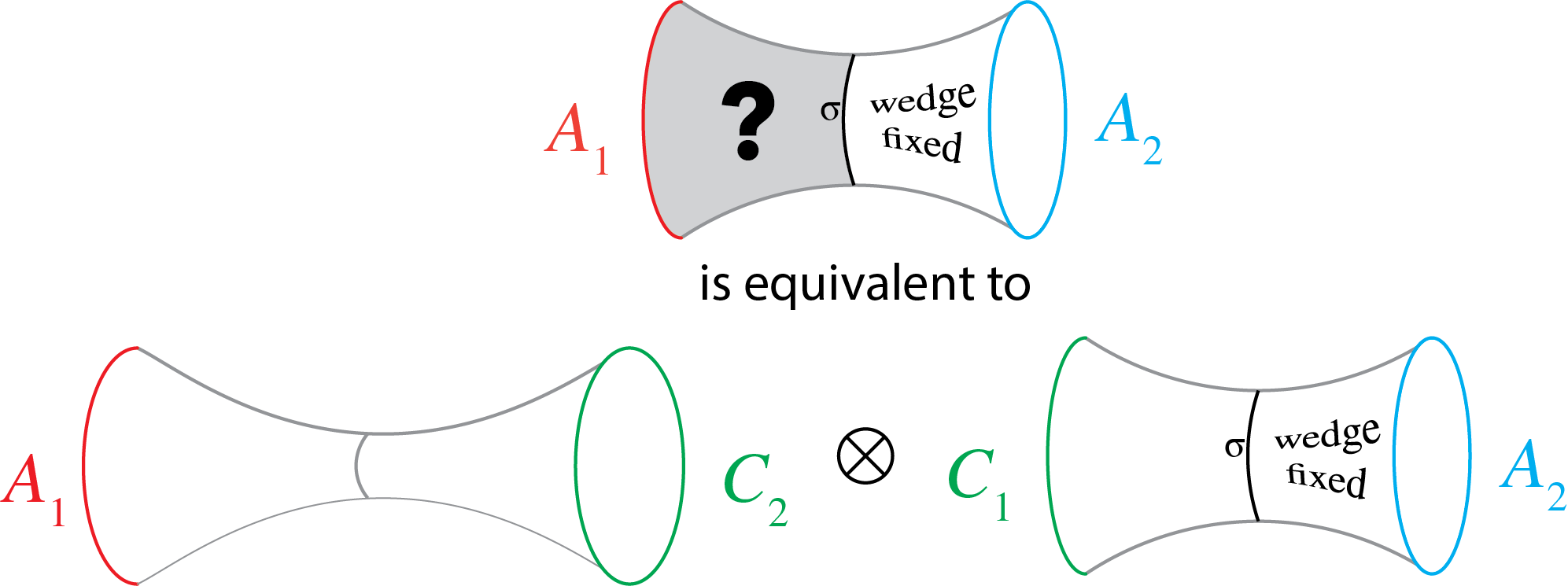}
  \caption{Holographic Shannon definition is defined almost identically to the HRT formula for covariant holographic entanglement entropy.
  In the figure on the left above, the area of the black surface gives the usual von Neumann entropy of the red region $A_1$.  In the figure on the
  right, the area of the black surface, which now includes $A_1$ as well as its entangling surface, produces what we call holographic Shannon entropy.}
  \label{fig:equivalence}
\end{figure}

The subtleties of our definition of equivalence are well-illustrated by an example shown in figure \ref{fig:equivalence}.  An AdS-Schwarzschild
black hole with two asymptotic regions $A_1$ and $A_2$ is shown in the upper figure.   The surface $\sigma$ is the entangling surface of $A_1$ in
this example, but that is unnecessary.  The lower half of the figure is the disjoint union of two spacetimes. (We use the tensor product symbol
as a reminder of how such a spacetime is treated in AdS/CFT).  The lower pair is equivalent to the original black hole on the $A_2$ side
of $\sigma$.  Notice how the new spacetime still has copies of the original regions $A_1$ and $A_2$, but that it also has additional
new asymptotic regions.  Also note that one of the black holes in the new spacetime has a different structure.

There is a critical observation to be made about this definition of equivalence.  Suppose that $\bar{M} \in \mathcal{C}(M_0,\sigma, A_2)$.
Then there must be a surface $\bar{\sigma}$ in $\bar{M}$ which is to homologous to $\bar{A}_2$ and which has the same (regulated)
area as $\sigma$.  However, \emph{there is no guarantee that $\bar{\sigma}$ is also homologous to $\bar{A}_1$}.  This is because
the spacetime $\bar{M}$ may have a conformal boundary with additional regions beyond $\bar{A}_1$ and $\bar{A}_2$.

The size of $\mathcal{C}(M_0,\sigma, A_2)$ is a seemingly good starting place when assessing the entropy associated with not knowing the gravitational
information on the $A_1$ side of $\sigma$.  However, this set is far too large as it includes arbitrary and completely pointless extensions of the spacetime
with additional disconnected components.

To solve this problem, we follow \cite{outer_1,outer_2} and take $\bar{M} \in \mathcal{C}(M_0,\sigma, A_2)$ without restriction, but we compute the HRT
entropy only of $\bar{A}_1 \cup \bar{A}_2$ which, as mentioned above, is not necessarily the entire boundary.  This entropy is bounded
upon regularization to some maximal value $S_\star(A_1) + S_\star(A_2)$  which is proportional to the regulated CFT volume of $A_1$ and $A_2$.\footnote{There
 is no need to use bars on arguments to $S_\star$ since $A_i$ and $\bar{A}_i$ have the same volume (after regularization).}
 
 This estimate of the entropy associated with coarse-graining over the region on the $A_1$ side of $\sigma$  in the spacetime $M_0$ is what we refer to as geometrical outer entropy and is denoted as
 \begin{equation}
\overset{\circ}{S}(\sigma, A_2) = \sup \left\{  \frac{\abs{\gamma(\bar{B}(M))}} {4 G_{\mathrm N}} \:  \middle| \: M \in  \mathcal{C}(M_0,\sigma, A_2)  \right\}. 
 \end{equation}
 In this expression, $\bar{B}(M)$ means the region $\bar{A}_1 \cup \bar{A}_2$ associated with the spacetime $M$ and $\gamma(A)$ is the HRT surface of $A$.
 Note that it would be better to put $M_0$ as an argument for $\overset{\circ}{S}$ but we avoid this since it will be clear from context what initial spacetime we are working with.
 
 This formula for geometrical outer entropy is based on the same ideas as in section \ref{sec:motivation} and in \cite{outer_1}.  We are essentially imagining that our original spacetime corresponds to
 a pure quantum state $\psi_0$ on $A_1 \cup A_2$ and that $\overset{\circ}{S}$ is computing, by way of the HRT formula, the von Neumann entropy of the density matrix
 \[
 \rho = \frac{1}{n} \sum_{\psi \in \mathcal{C}(\psi_0, \sigma, A_2)} \ket{\psi} \bra{\psi}
 \]
where $n$ is a normalizing constant and $ \mathcal{C}(\psi_0, \sigma, A_2)$ is the collection of pure quantum states on $A_1 \cup A_2$ subject to the condition that the spacetime is fixed on the
$A_2$ side of $\sigma$.

We now consider the case where $\sigma = \gamma(A_1)$ so that the outer wedge $\W(\sigma, A_2)$ is in fact the entanglement wedge $\EW(A_2)$.  In this case we will refer
to $\W(\sigma, A_2)$ as the outer wedge of $A_1$ and denote it by $\OW(A_1)$.
The argument is now similar to that in section \ref{sec:motivation}.  $\overset{\circ}{S}(\sigma, A_2)$ is bounded above by $S_\star(A_1) + \frac{|\gamma(A_1)|}{4 G_{\mathrm N}}$ because
it is the supremum of the set 
\[
\left\{  \frac{\abs{\gamma(\bar{B}(M))}} {4 G_{\mathrm N}} \:  \middle| \: M \in  \mathcal{C}(M_0,\sigma, A_2)  \right\},
\]
any element of which is an HRT entropy of $\bar{A}_1 \cup \bar{A}_2$.  The HRT formula satisfies subadditivity and the entropy of $\bar{A}_2$ is guaranteed to be $\frac{|\gamma(A_1)|}{4 G_{\mathrm N}}$
by the definition of $\mathcal{C}$ while the entropy of $\bar{A}_1$ is bounded by $S_\star(A_1)$ upon regularization.

To see the the upper bound is saturated, we can let $\bar{M}$ be the disjoint union of two bulks $M_1$ and $M_2$.  $M_1$ is an isometric copy of $M_0$, but the regions corresponding to $A_1$ and $A_2$ will
be denoted by $\bar{C}_1$ and $\bar{A}_2$, respectively.  (This labeling is extremely important.). The second spacetime $M_2$ has a boundary $\partial M_2$ which also consists of copies of $A_1$ and $A_2$
but we label them by $\bar{A}_1$ and $\bar{C}_2$, respectively.  The spacetime $M_2$ is not the same as $M_0$.  Instead, we take it to be a black hole with the largest possible temperature, after regulation, so that entropies of subregions
go like regulated CFT volume.  In this spacetime, $\gamma(\bar{A}_1 \cup \bar{A}_2)$ is a disjoint union of a copy of (a copy of) the original HRT surface $\gamma(A_1)$ and a surface pressed up against $A_1$ at the boundary.
The outer entropy with $\OW(A_1) = \EW(A_2)$ held fixed is thus the sum of two terms: the holographic entanglement entropy of $A_1$ in the original spacetime, and a much larger term proportional to the regulated volume of $A_1$.

\section{Motivation and existence: stabilizer states}
\label{sec:stabilizer}
There is a very different motivation for the holographic Shannon entropy formula which comes a property of stabilizer quantum states.
An elegant discussion of stabilizer states can be found in \cite{stabilizer_cone}, but a detailed knowledge of the subject is not a prerequisite for this section.

Let $d \geq 2$ be an integer and let $\mathcal{H} = \left({\mathbf C}^d\right)^{\otimes n}$ be the Hilbert space of $n$ qu-$d$-its.  
Gross and Walter proved that if $\rho$ is a stabilizer state on $\mathcal{H}$, then there exists a classical joint probability distribution $p$ 
 with the following entropic property.
Suppose that $I$ is any subset of $\{1,2,\ldots,n\}$ and that  $\rho_I$ and $p_I$ denote, respectively, the reduced density matrix and marginal probability distribution
on the copies of  $\left({\mathbf C}^d\right)$ and $X_i$ indexed by $i \in I$.  Then, the von Neumann entropy $S(\rho_I)$ and Shannon entropy $H(p_I)$
are related by
\begin{equation}
\label{gross_walter}
S(\rho_I) = H(p_I) - |I| \log d.
\end{equation}
The factor of $\log d$ arises because we insist on using natural logarithms rather than base $d$ to avoid confusion involving holographic entanglement.

The result \eqref{gross_walter} establishes a remarkable relationship between classical and quantum information in certain cases.  The most direct consequence of the relationship
 is that every balanced classical information inequality\footnote
 {When we refer to classical information inequalities, we mean 
inequalities linear in Shannon entropies of probability distributions that are valid for all distributions.   In other words, let $\mathcal{I}$ be the power set of $\{1,\ldots, n\}$
with the empty set removed and fix a function $c:{\mathcal I} \to {\mathbf R}$.  Suppose that for any probability distribution $p$ on $n$ random variables, $\sum_{I \in {\mathcal I}} c(I) H(p_I) \geq 0$ where $p_I$ is the marginal of $p$ on the subset of variables indexed by $I$.
Then, we say that $c$ furnishes a classical information inequality.  Strong subadditivity is an example, but there are many more \cite{zhang_yeung}. 
 Information inequalities like $H(A) + H(B) \geq H(AB)$ are \emph{balanced} which means that for any $j \in \{1, \ldots , n\}$, $\sum_{I \in \{I' \big| i \in I'\}} c(I) = 0$.  Inequalities like $H(AB) \geq H(A)$ are not
balanced because $B$ appears only on the left hand side.
}
holds for von Neumann entropies of stabilizer quantum states \cite{stabilizer_classical, stabilizer_cone}.  This fact follows directly from \eqref{gross_walter} because the term proportional to $|I|$ cancels on both sides of a balanced inequality.\footnote{
There is an unsettled conjecture due to Cadney, Linden, and Winter (CLW) that this result is not a special property of stabilizer states but rather that every quantum state satisfies every balanced classical entropy inequality \cite{CLW}.  This is what happened with strong subadditivity \cite{doi:10.1063/1.1666274}, but given how much harder the proof of quantum strong subadditivity is than that of classical strong subadditivity, it could be extremely difficult to provide a similar proof for even the quantum version of the Zhang-Yeung inequality \cite{zhang_yeung}. There are holographic consequences if the CLW conjecture is correct \cite{Mm_not_enough}.}

Equation \eqref{gross_walter} already looks like the holographic Shannon entropy formula because upon regularization, the volume of a boundary region is proportional to the number of constituent subsystems.
In fact, we can now make a powerful assertion.  Consider a static bulk where the Ryu-Takayanagi formula applies.  We can divide the boundary into a large number of subregions $R_1, R_2, \ldots R_n$ and assign to each
a single qubit upon regularization.  The Ryu-Takayanagi formula allows us to compute the von Neumann entropy of $\cup_{i \in I} R_i$ for any subset of indices $I$.  These entropies can be reproduced to arbitrary
precision by stabilizer states on $n$ parties by an argument involving random tensor networks \cite{random_tensor}.  Thus, there is a joint probability distribution $p$ on $n$ random variables, one associated to each of the regions $R_i$,
such that the Shannon entropy $H(p_I)$ of the marginal associated with the region $\cup_{i \in I} R_i$ is given by the area of its RT surface plus $|I| \log 2$.  By making the standard holographic identification
\[
\log 2 = \frac{|R_i|}{4 G_{\mathrm N}}
\]
for every qubit-containing subregion $R_i$, we obtain equation \eqref{basic_def}.

So, up to the cumbersome issue of regularization, we can conclude that in static cases there is some probability distribution jointly defined on local random variables on the boundary with Shannon entropies
given by equation  \eqref{basic_def} for all marginals, no matter their topology.  
In dynamical cases where entangling surfaces are computed with the HRT formula or by the approach of \cite{SW},
we are not yet in a position to prove that a probability distribution realizing equation \eqref{basic_def} exists.  However,
such a distribution exists if the dynamical holographic entropy cone for $N$ parties lies within the $N$ party stabilizer cone \cite{stabilizer_cone}
as this would allow us to apply the result of \cite{stabilizer_classical} for the dynamical case just like the static case.

In fact, it may turn out that the dynamical case isn't nearly as threatening as it appears.  There is currently a great deal
of interest in the open problem asking whether or not the dynamical and static holographic entropy cones for $N$ parties are in
fact, the same thing!  The monogamy of mutual information inequality \cite{hol_mmi} guarantees that this is the case for $N \leq 4$ regions.
Unfortunately, the 5 party static entropy cone is defined by a collection of known inequalities \cite{hol_cone,Cuenca:2019uzx} of which
several cannot be proven with standard methods \cite{Mm_not_enough}.  These inequalities have already undergone some
dynamical testing \cite{Caginalp:2019mgu} with no violation thus far.  Recent investigations \cite{Hubeny:2018bri, Cui:2018dyq, Hubeny:2018trv, Hubeny:2018ijt}
have found structure to holographic entropies that holds in both the static and dynamical case.

\section{Properties of Holographic Shannon Entropy}

The conclusions above strongly suggest that we take equation \eqref{basic_def} seriously.  Even if the outer entropy argument has an element of hand-waving to it and even 
if the stabilizer state argument is taken only as an existence proof,
 there is still an accumulation of evidence that for some reason or another, \eqref{basic_def} is an important finding.
With this in mind, we now clean up the geometrical area formula, providing a more formal definition, and we quickly prove some of its properties that suggest a Shannon entropy interpretation.

Fix a globally hyperbolic spacetime $M$ (or at lease a spacetime with a globally hyperbolic conformal compactification) obeying the null energy and generic conditions.  \emph{Do not assume that $M$ is asymptotically locally AdS}.
Let $\sigma$ be a connected compact spacelike codimension 2 surface in $M$ with future-directed null orthogonal vector fields $k$ and $l$.  
 $\theta_k$ and $\theta_l$, satisfy $\theta_k \geq 0$ and $\theta_l  \leq 0$.  
 Take a Cauchy surface $\Sigma_0$ that contains $\sigma$ and consider the domain of dependence of the part of $\Sigma_0$ which
 is inside of $\sigma$.  By ``inside'' we mean in the direction of $l-k$ from $\sigma$.  Denote this domain of dependence by $D_\sigma$.  
 
  Let $A$ be a subregion of $\sigma$ and let $\gamma(A)$
  denote the smallest extremal surface homologous to $A$  that lies inside of $D_\sigma$.  The area of $\gamma(A)$ will still be called holographic entanglement entropy
  even though it is not anchored to an AdS boundary.  Like conventional holographic entanglement entropy \cite{hol_ssa, hol_mmi, Mm}, the construction here satisfies monogamy of mutual information and strong subadditivity \cite{SW}.

The \emph{holographic Shannon entropy} of $A \subset \sigma$ is the quantity
\[
H(A) = \frac{1}{4 G_{\mathrm N}} \left( |A| + |\gamma(A)| \right)
\]
where $|A|$ is the area of $A$ and $\gamma(A)$ is the entangling surface of $A$ as defined in the previous paragraph.  Note that in this context, the neither term is divergent, although
the first term is always greater than or equal to the second \cite{SW}.

The most important geometrical results we currently know about this construction are the following pair of theorems:

\begin{theorem}
Let $A$ and $B$ be subregions of $\sigma$ which are disjoint except, perhaps, at their boundaries. Then, $H(AB) \geq H(A)$.
\proof 
Recall that the entangling surfaces of regions on $\sigma$ lie in the domain of dependence $D_\sigma$.
Wall's entanglement wedge nesting theorem \cite{Mm} was generalized in \cite{SW} to show that we can find a topologically closed spacelike surface $\Sigma$ anchored to $\sigma$
such that
\begin{enumerate}
	\item The domain of dependence of $\Sigma$ is $D_\sigma$
	\item $\gamma(A) \subset \Sigma$ and $\gamma(AB) \subset \Sigma$
	\item $\gamma(A)$ and $\gamma(AB)$ are minimal area surfaces on $\Sigma$ anchored to $\partial A$ and $\partial (AB)$ respectively.
\end{enumerate}
Notice that $\gamma(AB) \cup B$ is a non-minimal surface on $\Sigma$ anchored and homologous to $A$.
Thus, $|\gamma(A)| \leq |\gamma(AB)| + |B|$ so
\[
|A| + |\gamma(A)| \leq |A| + |B| + |\gamma(AB)|.
\]\\
\end{theorem}

\begin{theorem}
Every balanced inequality that applies to covariant holographic entanglement on compact a compact surface applies holographic Shannon entropy.
\proof This is obvious because in balanced inequalities, the term proportional to volume of a region cancels on both sides of the inequality.
\end{theorem}

The first of these theorems relates to the most famous distinction between classical and quantum correlation.  Consider 
two random variables $A$ and $B$ and their joint probability distribution $p$.  The entropy of $A$ is the expectation value
of $-\log p(A)$ so it measures the expected ``surprisal'' of $A$.  If we always measure $B$ before measuring $A$, we should
expect to be less surprised by $A$, so the conditional entropy $H(A | B)$, defined as the expectation value of $-\log p(A | B)$, must be no greater than $H(A)$.
Because $p(a,b) = p(a | b) p(b) $, it follows that  $ H(AB) = H(A | B) +   H(B)$, and from this we see that $H(A | B) \leq H(A)$ is the statement of subadditivity,
a property that classical and quantum correlations hold in common.  However, quantum mechanically we can have $S(AB) = 0$ while $S(A) > 0$
so that $S(A | B) < 0$ while classically $H(A | B)$ is the expectation value of a positive random variable.  

Shannon entropy thus satisfies monotonicity: $H(AB) \geq H(A)$ while von Neumann entropy does not.  The holographic Shannon entropy
formula and the HRT formula share this critical distinction.  

On the other hand, the second theorem shows that there is a large family of information
inequalities that the two constructions have in common.  In particular, we immediately conclude that $H$ satisfies monogamy of mutual information and
strong subadditivity, the first of which is not satisfied by all probability distributions.  The second theorem also allows us to conclude, becuase
of the findings of \cite{Mm_not_enough}, that $H$ satisfies a large class of known classical inequalities.  
These include those of \cite{zhang_yeung, Makarychev2002ANC, Matus}.  However, the balanced inequalities found in \cite{zhang_yang}
are not yet proven for the dynamical holographic entropy cone \cite{Mm_not_enough} and are thus not proven for holographic Shannon entropy.

\section{Remaining Questions}
The simple construction given here is promising and problematic. In the static case with a compact or regulated AdS boundary, there is no doubt
that the holographic Shannon entropy formula is really computing all of the Shannon entropies of the marginals of some probability distribution.
In the general dynamical case, there are already enough known inequalities satisfied by $H$, especially monotonicity, to have
some confidence that there is no need to restrict to static spacetimes.
But what probability distribution are we talking about?  The stabilizer state methodology can be taken as an existence proof, but it leaves much to be desired.
Is there some classical probability distribution defined on local subsystems of the boundary that arises for a \emph{physical} or natural reason?
Does this hypothetical distribution directly relate to the concepts of outer entropy of entangling surfaces?
Is there a classical form of holography that could have an impact on classical information theory?  Some steps in this direction
appear to have been considered in \cite{PhysRevD.96.066005}.

The most important question about this mysterious probability distribution is its relationship to outer entropy.  Counting gravitational states is an
old and challenging problem in quantum gravity and it has been largely hindered by the difficulty of defining local subsystems in general relativity.
However, entanglement wedge reconstruction offers one way that spatial regions can be separated from one another, and it is obviously important to ask
how many different spacetimes or states fit into an entanglement wedge.  The answer is apparently the obvious holographic one: the area of boundary of the wedge including
both the bulk and boundary sides of the wedge.  The Bousso entropy bound is saturated \cite{Nomura:2013lia}.

\section*{Acknowledgments}
I would like to thank Donald Marolf, Massimiliano Rota, Fabio Sanches, and Aron Wall for important discussions.
This work was supported by the generosity of the Len DeBenedictis Fellowowship and by additional support from the
University of California.

\bibliography{bib} 
\bibliographystyle{ieeetr}

\end{document}